\documentclass[]{spie}  

\usepackage{xcolor}

\usepackage{amsmath,amsfonts,amssymb}
\usepackage{graphicx}
\usepackage[colorlinks=true, allcolors=blue]{hyperref}
\usepackage{multirow}

\title{SAXO+, the second-stage adaptive optics for SPHERE: NCPA compensation and dark-hole loop with a pyramid wavefront sensor}

\author[a]{Johan Mazoyer}
\author[a]{Charles Goulas}
\author[a]{Raphaël Galicher} 
\author[a]{Axel Potier}
\author[a]{Fabrice Vidal} 
\author[a]{Florian Ferreira}
\author[a]{Arnaud Sevin}

\author[b]{Clémentine Béchet}
\author[c]{Isaac Bernardino Dinis}
\author[a]{Anthony Boccaletti} 
\author[d]{Gael Chauvin} 
\author[e]{Fausto Cortecchia} 
\author[e]{Emiliano Diolaiti}
\author[f]{Nicolas Galland} 
\author[f]{Caroline Kulcsár} 
\author[b]{Maud Langlois}
\author[e]{Matteo Lombini}
\author[g]{Julien Milli}
\author[h]{Mamadou N’diaye}
\author[f]{Henri-Fran\c{c}ois Raynaud}
\author[e]{Laura Schreiber}
\author[b]{Michel Tallon}
\author[i]{Arthur Vigan}
\author[c]{Fran\c{c}ois Wildi}

\affil[a]{LIRA, Observatoire de Paris, Université PSL, Sorbonne Université, Université Paris Cité, CY Cergy Paris Université, CNRS, 92195 Meudon, France}

\affil[b]{Université Lyon 1, ENS de Lyon, CNRS, CRAL, UMR 5574, Saint-Genis-Laval, France} 
\affil[c]{Départment d’astronomie de l’Université de Genève, 51 ch. des Maillettes Sauverny, 1290 Versoix, Switzerland} 

\affil[d]{MPIA, Max-Planck-Institut f\"ur Astronomie, K\"onigstuhl 17, D-69117 Heidelberg, Germany}

\affil[e]{INAF Osservatorio di Astrofisica e Scienza dello Spazio di Bologna, Via P. Gobetti 93/3, 40129, Bologna, Italy} 

\affil[f]{Université Paris-Saclay, Institut d'Optique Graduate School, CNRS,  Laboratoire Charles Fabry, 91127 Palaiseau, France}
\affil[g]{Université Grenoble Alpes, CNRS, IPAG, 38000, Grenoble, France} 
\affil[h]{Université Côte d’Azur, Observatoire de la Côte d’Azur, CNRS, Laboratoire Lagrange, France} 
\affil[i]{Aix-Marseille Univ., CNRS, CNES, LAM, Marseille, France}

\authorinfo{Send correspondence to johan.mazoyer@obspm.fr}

\begin{document} 
\maketitle

\begin{abstract}
The SAXO\texttt{+} upgrade of the VLT/SPHERE adaptive optics system introduces a second-stage near-infrared pyramid wavefront sensor to improve high-contrast imaging, making accurate calibration of non-common path aberrations (NCPAs) essential to fully exploit its performance. This work refines the expected level of NCPAs in SAXO\texttt{+} and presents the calibration procedures developed for static NCPA compensation and focal-plane dark-hole control. Monte Carlo simulations based on an updated Zemax optical model were used to estimate the NCPA error budget. These simulations are in good agreement with previous measurements on SPHERE and with the assumptions adopted in earlier performance studies. We also propose a calibration strategy that offloads most static aberration correction to the first-stage deformable mirror while preserving the second-stage mirror stroke for high-speed adaptive optics correction. These results validate the expected SAXO\texttt{+} optical quality and establish the calibration framework required for efficient NCPA compensation and focal-plane wavefront control during future on-sky operations.

\end{abstract}

\section{INTRODUCTION: THE SAXO\texttt{+} INSTRUMENT}
\label{sec:intro}

The main goal of direct imaging is to visually capture the environments surrounding stars, from exoplanets to debris and protoplanetary disks. Through high-contrast imaging, it becomes possible to locate objects near stars with precision and study the light they emit or reflect. In the case of exoplanets, this approach helps determine their orbits and the physical and chemical properties of their atmospheres.

The difficulty of this approach lies in the enormous brightness gap (spanning from $10^{-4}$ to $10^{-10}$)and the minimal angular distances (from a fraction of an arcsecond to a few arcseconds for the closest stars) between a planet and its star\cite{galicher_imaging_2024}. To date, fewer than 1\% of the 6,000 confirmed exoplanets have been directly imaged. The newest high-contrast instruments, including SPHERE \cite{beuzit2019_SPHEREExoplanetImager}, GPI \cite{macintosh2014_FirstLightGemini}, SCEXAO \cite{lozi2020_StatusSCExAOInstrument}, and MagAO-x \cite{Males2018SPIE_MagAO-X}, have successfully identified planetary-mass companions orbiting nearby stars. By combining extreme adaptive optics, coronagraphy, and sophisticated calibration and post-processing, these systems effectively block starlight to uncover circumstellar objects at separations of just a few resolution elements.

The current capabilities of SPHERE \cite{Squicciarini_SHINE_2026} remain insufficient for imaging planets at small angular separations that were identified through alternative methods such as astrometry, radial velocity, or microlensing. Thus, further improvements in contrast at these separations are necessary to enable a comprehensive analysis of planet distribution across the full range of orbital distances. Current coronagraphs' performance is significantly degraded by atmospheric turbulence and optical manufacturing defects. While adaptive optics (AO) systems can correct for atmospheric distortions, they cannot address non-common path aberrations (NCPAs)—static or quasi-static aberrations that differ between the wavefront sensing path and the astrophysical path. Since the AO system does not perceive the same aberrations as the coronagraph, it fails to correct them precisely. NCPAs introduce a speckle pattern in the coronagraph image, which evolves slowly over time due to thermal variations, mechanical flexure, or other factors, often mimicking the signal of an exoplanet.

\begin{figure} [ht]
\begin{center}
\begin{tabular}{c} 
\includegraphics[width=0.8\linewidth]{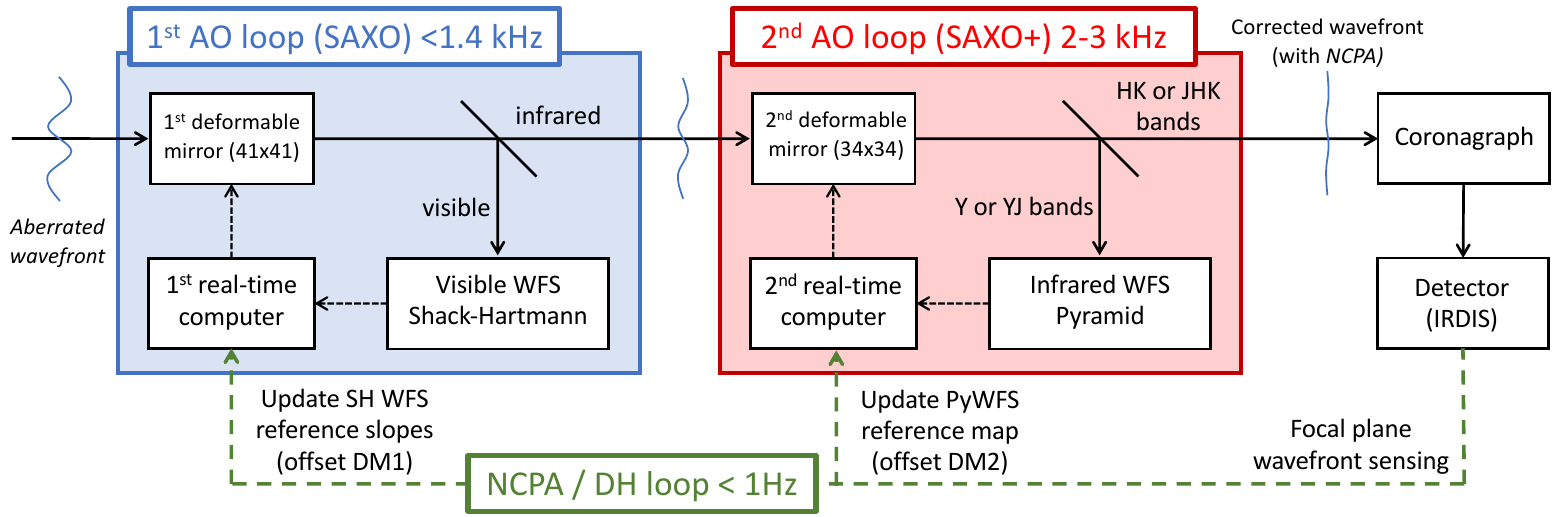}
\end{tabular}
\end{center}
\caption[]{\label{fig:saxo_3loops} \textbf{SAXO\texttt{+} outline}. In blue: current SAXO system. In red: second loop of SAXO\texttt{+}. In green: NCPA correction (adapted from Goulas et al. 2026\cite{goulas_non-common_2026}).}
   \end{figure}

For bright targets under favorable conditions (when atmospheric turbulence is well-compensated by the AO system), SPHERE coronagraph images are generally dominated by NCPA-induced speckles and methods have been developed to actively correct these speckles during observations \cite{vigan2019_CalibrationQuasistaticAberrations,potier_increasing_2022}. However, the AO system is often restricted by the sensitivity of the wavefront sensor and temporal errors in the control loop~\cite{cantalloube_sphere_2019}. For faint stars and/or bad condictions, the current AO performance therefore limit the measurement and calibration of the NCPA-induced speckles. To address these issues and enhance detection capabilities, SAXO is being developed into SAXO\texttt{+}\cite{boccaletti_sphereplus_2020}. As depicted in Fig.\ref{fig:saxo_3loops}, SAXO\texttt{+} introduces a secondary AO loop downstream of the original SAXO stage. The first stage operates at 1.38 kHz, utilizing a 41×41 Shack-Hartmann (SH) WFS in visible light alongside a 41×41 DM. The second stage, designed to correct SAXO’s temporal errors, operates at up to 3 kHz and employs a near-infrared (1.2 µm) 50×50 pyramid WFS (pyWFS) \cite{Ragazzoni_1996}, more sensitive than the SH WFS. This stage incorporates a 34×34 kilo-DM from Boston Micromachines, although only 28 actuators are active across the pupil diameter. The two loops are controlled by separate real-time computers, and this analysis assumes the loops function independently. With a 2 loop system SAXO\texttt{+} coronagraphic images are expected to be predominantly speckle-limited for bright stars, making a good correction of NCPAs all the more necessary.  

NCPAs needs to be measured at the focal plane mask level or even after to prevent differential aberrations, using directly the Science detector (IRDIS), and corrected in a third loop, as depicted in Fig.\ref{fig:saxo_3loops}. NCPA compensation in AO systems is typically implemented by introducing an offset in WFS  signals such (e.g. SH reference slopes), to counteract aberrations. This approach is highly effective when the WFS operates within its linear range, as is the case for SH WFS. However, for a pyWFS, the linear range is significantly more limited. Even with a perfect estimation of NCPAs, the accuracy of applying an offset to the normalized pixels on the PyWFS detector (hereafter referred to as the PyWFS response) diminishes as NCPA levels increase and as the incoming wavefront degrades (i.e. under bad atmospheric conditions). Consequently, only a portion of the NCPAs may be corrected, and in extreme cases, the correction could even degrade performance. To mitigate this limitation, estimating the optical gains of the NCPA modes can help extend the linear range and improve correction accuracy.\cite{korkiakoski2008_ImprovingPerformancePyramid,deo2019_TelescopereadyApproachModala,esposito2020_OnskyCorrectionNoncommon,chambouleyron2020_PyramidWavefrontSensor}.

In SAXO, NCPA were measured using a Zernike wavefront sensor (ZWFS)\cite{ndiaye2016_ApodizedPupilLyot}. This process was iterated in a closed NCPA correction loop while the AO loop remained active, using internal source and on sky\cite{Vigan_zelda_2019}. Zelda is currently running everyday to monitor NCPA evolution. In the SAXO\texttt{+} system, Zelda will be used as a daily NCPA calibration procedure. While this technique effectively minimizes phase aberrations in the pupil plane—thereby reducing stellar speckle intensity in the coronagraph image—it does not address starlight residuals caused by amplitude aberrations or diffraction effects from telescope spiders. To suppress these residuals, a dark-hole loop is employed. This loop aims to minimize quasi-static intensity in a specific region of the focal plane. In SPHERE, \cite{potier_increasing_2022} estimated the speckle electric field in the IRDIS image using the Pair-Wise Probing (PWP) technique \cite{giveon_closed_2007}. PWP records four coronagraph images by actuating the DM actuators with a known amplitude, modulating the speckle intensity over time. Once the electric field is known, the Electric Field Conjugation (EFC) method is used to minimize starlight intensity in a targeted area of the coronagraph image by optimizing the DM shape. Typically, the dark hole loop requires multiple iterations of sensing and minimization due to linear assumptions or errors in the light-propagation model. Both the speckle modulation for sensing with PWP and the speckles' correction with EFC require precise of movement of the DM, which can only be achieved in a close loop by offsetting the WFS reference image. Both ZWFS and Dark-hole method were successfully demonstrated with an SH WFS on SPHERE, on an internal source and on-sky \cite{vigan2022_CalibrationQuasistaticAberrations, potier_increasing_2022}. However, this modification with the pyWFS is \textit{a priori} much harder as this WFS response is not linear and depend on the incoming WFS condition. The correction allowed with these methods were therefore very dependent on the actual amount of NCPA present on the system and also on the quality of the wavefront received from the pyramid. For this reason, we investigate optical modal gains to account for the changeing WFS conditions.  

Most of the results presented at this conference were recently published in Astronomy \& Astrophysics (Goulas et al. 2026 \cite{goulas_non-common_2026}). To avoid reprinting published results, this proceeding will only quickly recall the most important results of this paper and describe in more details expected level of NCPA on SPHERE+ and calibration procedure. 

\section{Brief summary of Goulas et al. (2026)}

Goulas et al. (2026)\cite{goulas_non-common_2026} investigates the compensation of NCPAs and the implementation of a focal-plane dark-hole control loop in the context of the SAXO\texttt{+} adaptive optics upgrade for SPHERE. Using end-to-end numerical simulations, Goulas et al. (2026) evaluate the performance of a near-infrared pyramid WFS operating downstream of the primary AO system under a range of observing conditions. 

We first described an on-sky calibration method for the pyWFS modal optical gains, which accounts for the sensor's nonlinear response under realistic atmospheric turbulence, adapted from Esposito et al. 2020 \cite{esposito2020_OnskyCorrectionNoncommon}. 

We then showed that static NCPA compensation can reduce the residual stellar light in coronagraphic images by up to a factor of 20 for bright targets observed under typical seeing conditions. We also demonstrate that a closed-loop dark-hole algorithm, based on iterative focal-plane wavefront control, provides a substantially larger gain than static NCPA compensation alone. The dark-hole loop suppresses residual starlight by up to a factor of 200 by directly minimizing the intensity of quasi-static speckles in the science image, including contributions from both phase and amplitude aberrations. The resulting parametric study establishes practical operating conditions—including loop gains, modulation radius, and observing regimes—for efficient implementation of focal-plane wavefront control in SAXO\texttt{+}.

Interestingly, while optical-gain calibration is beneficial for a single-stage pyramid AO system, it offers little additional improvement for the full SAXO\texttt{+} architecture, whose dual-stage AO design is intrinsically more robust to pyWFS nonlinearities. Indeed, correction by the first stage ensure that the pyWFS mostly sees a high Strehl PSF and therefore operate in the linear regime. We show that the compensation of the pyWFS optical gains is useful to gain a factor of up to 2 in contrast in the coronagraph image for seeings above 1.5'' but has a negligible impact for better seeings.

All the estimation made on this study assumed 26 actuators across the telescope pupil diameter. However, since this paper, the Boston Deformable mirror actually procured for SAXO\texttt{+} is a BMC Kilo-C-3.5 with a total actuator count of 952 (34 across the full DM diameter, with corners not actuated) and a pitch of 400 $\mu m$. The projected pupil size on the second DM is 11.25mm, corresponding instead to 28.1 actuators  across the pupil diameter. We are currently adapting our simulation to the new DM. However, as the NCPA level, pyWFS characteristic, and AO performance of the first stage remain completely unchanged, we do not expect that this will change significantly the results. 

\section{Estimation of NCPA based on SAXO\texttt{+} optical model}

In Mazoyer et al. (2024) \cite{mazoyer_upgrading_2024}, we evaluated the amount of NCPA present in SAXO plus based on previous analysis made on SAXO with the ZWFS \cite{vigan2022_CalibrationQuasistaticAberrations}. The level of aberrations extracted were typically between 55 and 60 nm RMS, including  45-50 nm RMS in the SAXO correction zone ($<20\lambda/D$) and 40-45 nm RMS in the SAXO\texttt{+} correction zone ($<13 \lambda/D$). 
For SAXO\texttt{+}, we initially assumed that, as the number of optics outside of the common path was almost similar, the level of aberrations and frequency content after SAXO\texttt{+} will be comparable to these levels. For this reason, phase aberrations with similar level and frequency distribution were used by Goulas et al. (2026)\cite{goulas_non-common_2026}.

\begin{figure} [ht]
\begin{center}
\begin{tabular}{c} 
\includegraphics[width=0.8\linewidth]{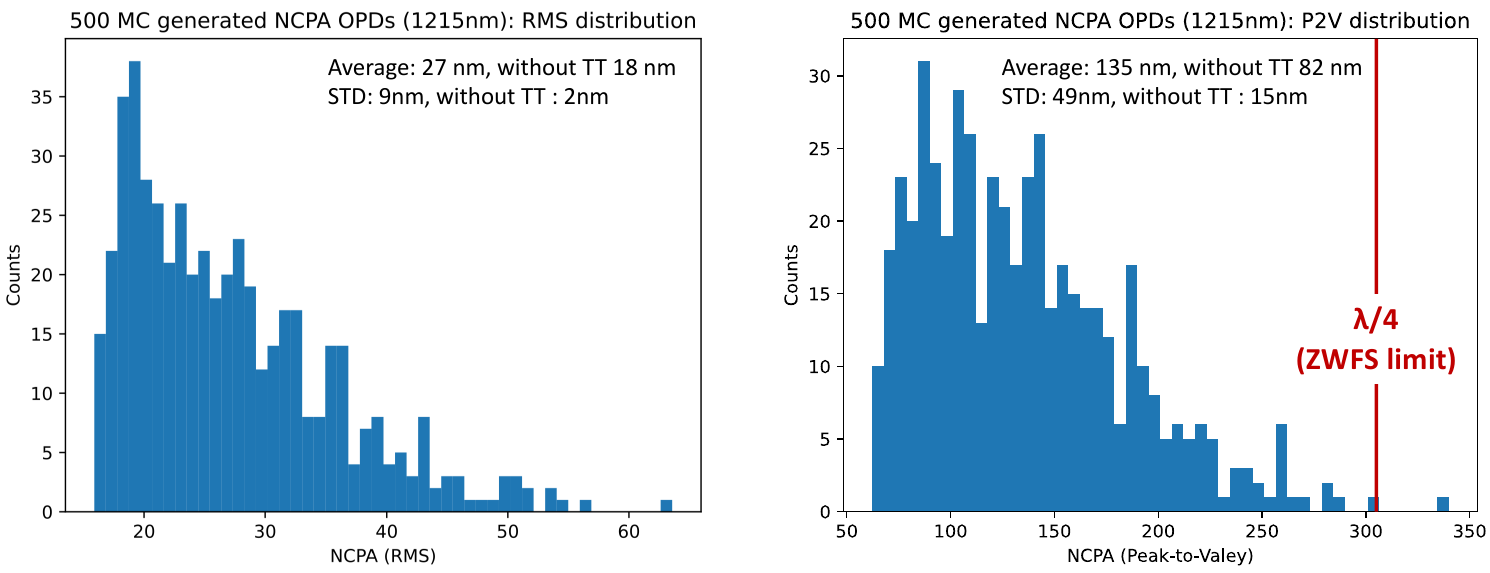}
\end{tabular}
\end{center}
\caption[]{\label{fig:NCPA_RMS_P2V}Distribution of NCPA level in RMS (left) and peak-2-valley (right) randomly generated with SAXO\texttt{+} optical model.}
\end{figure} 

In this section we refined this measurement using the expected alignment accuracy of SAXO\texttt{+}. Using the optimized Zemax OpticStudio® SAXO\texttt{+} optical model, a Monte-Carlo algorithms was used to generate optical path difference (OPD) maps at different point of the system, by varying each optics, shape, alignment, position in SAXO\texttt{+} around their optimum. In particular, we generated 500 OPD maps at focal plane of PyWFS and at the exit focal point of the SAXO\texttt{+} system. Assuming that the PyWFS corrects completely the static aberrations introduced in its channel, the difference of those 2 maps is the level of NCPA expected in SAXO\texttt{+} at the coronagraph focal plane level, due to optics shape and misalignments. The DTTS beam splitter was not included in this optical model as it is not a SAXO\texttt{+} optics, but is located very close to the focal plane so contribute very little to the total phase error budget. 

We plot the distribution of those NCPA OPDs, in RMS and Peak to valley (P2V) in Figure~\ref{fig:NCPA_RMS_P2V} showing that the level of expected NCPA is $27 \pm8$ nm RMS ($19 \pm2$ nm, if we exclude the TT that will mostly be corrected by DDTS) and $135 \pm9$ nm P2V in average ($82 \pm15$ nm, if we exclude the TT). The main results are that the expected level of aberrations are :
\begin{itemize}
    \item comparable to the ones measured on SAXO currently and used in our simulation so far,
    \item 99.5\% of the time below the cutoff wavelength of the ZWFS ($\lambda/4 \sim 300$nm in the J band) that will be used to measure the NCPA, which means we do not expect any folding in the measurement of NCPA.
\end{itemize}

\begin{figure} [ht]
\begin{center}
\begin{tabular}{c} 
\includegraphics[width=0.4\linewidth]{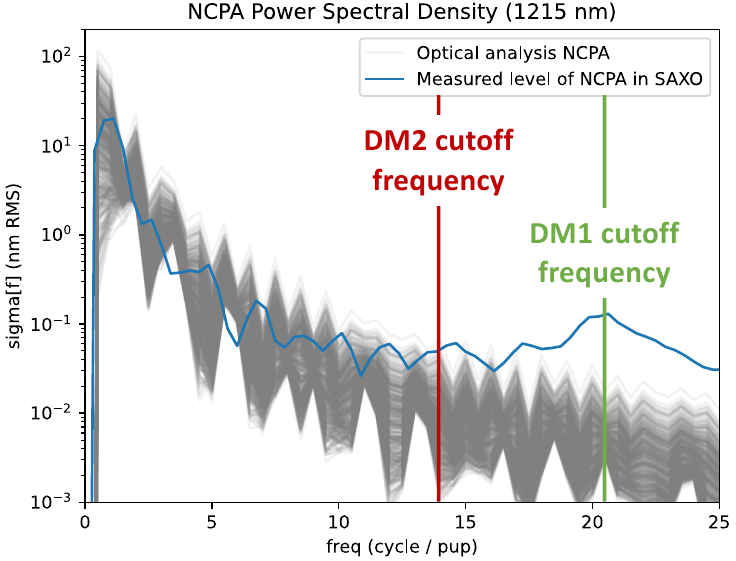}
\end{tabular}
\end{center}
\caption[]{\label{fig:ncpa_PSD_compar_Zelda} Power spectral density of generated NCPA OPD maps (in grey) compared to current level measured in SAXO (in blue).}
\end{figure} 

A comparison of the power spectral density of the generated NCPA OPD maps (in grey) compared to measured OPD maps calculated on SPHERE using Zelda (in blue) are shown on Figure~\ref{fig:ncpa_PSD_compar_Zelda} shows clear agreement for frequencies $< 12 \lambda /D$. Frequencies $< 12 \lambda /D$ are under evaluated with the current analysis. The cause are probably that the Zemax level do not take into acount surface rugosity. Finally, at large frequency (superior to the DM cutoff frequencies), the DM surface themselves introduce large amount or aberrations, not taken into account in this simulation and not correctable with our system. This quick study validates our assumption that the level of aberrations and frequency content after SAXO\texttt{+} will be comparable to the ones measured on SAXO.

\section{Calibration procedures}
In this section, we describe NCPA calibration and on-sky loops. SAXO\texttt{+} DM is a Boston Micromachine MEMS DM. MEMS deformable mirrors provide a smaller stroke than conventional piezoelectric devices (like the ones included in SAXO). In most observation cases, this is not a critical limitation for SAXO\texttt{+}, since the large-amplitude atmospheric aberrations are already compensated by the upstream first stage SAXO DM (DM1 usually called High-order DM, HODM). The second-stage mirror (DM1 usually called XPDM) therefore corrects only the residual high-order wavefront errors and benefits from the high temporal bandwidth offered by MEMS. There is one major exception, for very red stars (Mag G $>$ 14), when the first stage visible SHWFS car barely function and when the second stage is doing the heavy lifting. This aspect was studied in details in simulation \cite{Langlois2026_SAXOplussimu, tallon2026_SAXOpluscontrol}.

\begin{figure} [ht]
\begin{center}
\begin{tabular}{c} 
\includegraphics[width=0.8\linewidth]{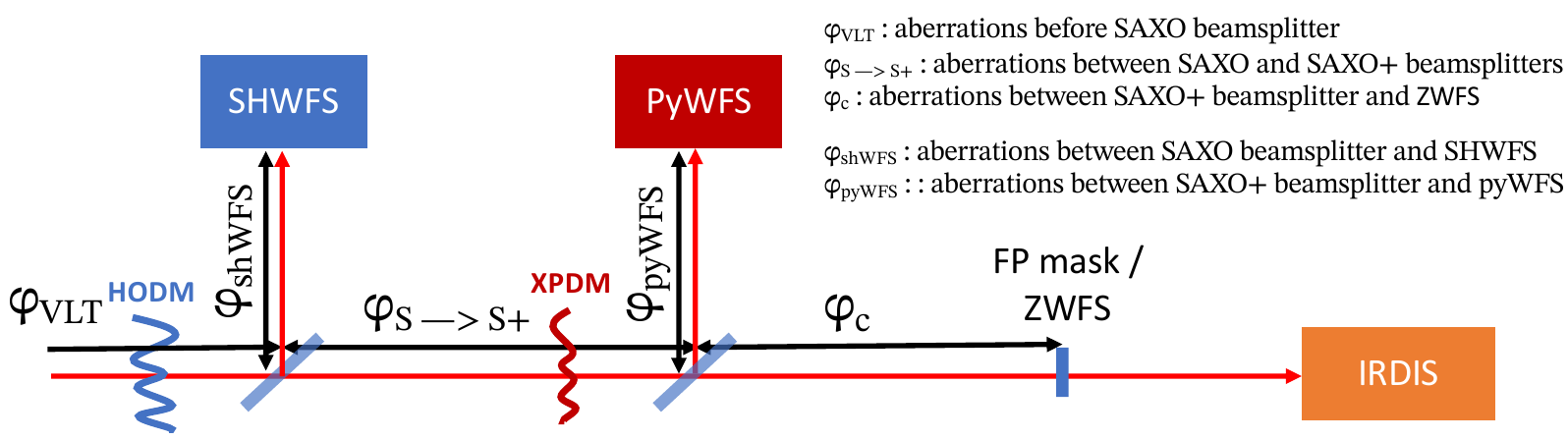}
\end{tabular}
\end{center}
\caption[]{\label{fig:ncpa_saxoplus} Simplified schematic of SAXO / SAXO\texttt{+} system describing all source of phase aberrations introduced by the optics, before SAXO ($\phi_{\text{VLT}}$), between SAXO and SAXO\texttt{+} ($\phi_{S \rightarrow S^+}$) and between SAXO\texttt{+} and the focal plane mask ($\phi_c$). We also introduce  phase aberrations introduced by the optics in each of the WFS arms ($\phi_{\text{SHWFS}}$ and $\phi_{\text{pyWFS}}$).}
\end{figure}

For NCPA this is not a major problem as we are moving in opposite direction : NCPA are mostly a nuisance for bright stars in good conditions, when the AO residuals in IRDIS focal plane are vey low and when strokes on the second DM are not important. For very red stars and/or in bad conditions, AO residuals dominates the image and NCPA correction becomes optional. Nonetheless, we designed a procedure to mostly offload NCPA on the SAXO DM with more strokes, instead of the second DM. Fig~\ref{fig:ncpa_saxoplus} shows a simplified schematic of SAXO / SAXO\texttt{+} system describing all source of non-common path aberrations and introducing notations. The goal of this procedure is to understand how to offset SHWFS slopes and pyWFS responses so that phase at the level of the coronagraph is annulled using only strokes on the first DM. 

\noindent \textbf{Step 1: Close SAXO loop --} With internal source, close loop on SAXO loop with no offset.
\begin{itemize} 
\setlength\itemsep{-0.1em} 
    \item \textbf{SHWFS} sees $0$
    \item \textbf{pyWFS} sees $\phi_{S \rightarrow S^+} - \phi_{\text{SHWFS}} + \phi_{\text{pyWFS}}$
    \item \textbf{ZWFS} sees $\phi_{S \rightarrow S^+} - \phi_{\text{SHWFS}} + \phi_c$\\
    \item \textbf{DM1 (HODM)} stroke: $- (\phi_{\text{VLT}} + \phi_{\text{SHWFS}})$
    \item \textbf{DM2 (XPDM)} stroke: $0$
\end{itemize}

\noindent \textbf{Step 2: ZWFS loop on SAXO --} Run NCPA correction loop with ZWFS, offsetting SHWFS iteratively (with SAXO\texttt{+} loop open).
\begin{itemize} 
\setlength\itemsep{-0.1em} 
    \item \textbf{SHWFS} is offested by $-(\phi_{S \rightarrow S^+} - \phi_{\text{SHWFS}} + \phi_c)$ / sees 0
    \item \textbf{pyWFS} sees $\phi_{\text{pyWFS}} - \phi_c$
    \item \textbf{ZWFS} sees $0$\\
    \item \textbf{DM1 (HODM)} stroke: $-(\phi_{\text{VLT}} + \phi_{S \rightarrow S^+} + \phi_c)$
    \item \textbf{DM2 (XPDM)} stroke: $0$
\end{itemize}

\noindent \textbf{Step 3: Offset pyWFS and close SAXO\texttt{+} loop --} Measure phase with pyWFS and offset pyWFS by this value. Close SAXO\texttt{+} loop.
\begin{itemize} 
\setlength\itemsep{-0.1em} 
    \item \textbf{SHWFS} is offested by $-(\phi_{S \rightarrow S^+} - \phi_{\text{SHWFS}} + \phi_c)$ / sees 0
    \item \textbf{pyWFS} is offested by $-(\phi_{\text{pyWFS}} - \phi_c)$ / sees 0
    \item \textbf{ZWFS} sees $0$\\
    \item \textbf{DM1 (HODM)} stroke: $-(\phi_{\text{VLT}} + \phi_{S \rightarrow S^+} + \phi_c)$
    \item \textbf{DM2 (XPDM)} stroke: $0$
\end{itemize}

\noindent Vigan et al. 2019 \cite{vigan2019_CalibrationQuasistaticAberrations} report around 20 nm RMS differences when applying internal source ncpa calibration directly on-sky. Potier et al. 2022 (SPIE) \cite{Potier2022spie} agrees that dark-hole dug on internal source shows poor performance when applyied directly on sky. For ZWFS performance change, these are likely due to NCPA aberrations evolving overtime due to moving optics or internal turbulence. These effects should be minimized in SAXO\texttt{+} \cite{mazoyer_upgrading_2024} but probably not completely cancelled. For Dark-Hole, the reason is that, in addition to phase aberrations evolution cited earlier, amplitude aberrations in the VLT optics cannot be correctly measured using the internal source. We therefore need to be able to perform NCPA and Dark-Hole correction directly on-sky to correct potential residuals. 

The routine described earlier runs most efficiently in internal source. Indeed, in sky, focal plane measurement with the ZWFS described in step 2 (and similarly dark hole loop) is heavily dependent on the amount wavefront residuals, which limit the measurement and therefore correction of NCPA and amplitude aberrations. For example, Vigan et al. 2019 : \textit{``we consider that the loss in sensitivity on sky is well understood and is directly related to the Strehl ratio of the observations''}. For this reason, if we really want to improve NCPA sensing and correction on-sky we need to be able to perform it with both loop closed, i.e. by offseting the pyWFS only. For this reason, we are also planning to have dark-hole loop and ZWFS loop direclty on SAXO\texttt{+} pyWFS. The exact amount of strokes that will be required on DM2 in that situation still need to be precisely simulated.

\section{Conclusion}

SAXO\texttt{+} enhances high-contrast imaging by addressing NCPAs through a secondary AO loop and advanced calibration. Refined NCPA estimates for SAXO\texttt{+} ($27 \pm 8$ nm RMS) align with SAXO measurements, validating our early assumptions. Advanced end-to-end simulations shows that pyWFS operating downstream of the primary AO systemis intrinsically more robust to pyWFS nonlinearities. Static NCPA compensation and dark-hole algorithms greatly improve performance of the algorithm. We presented internal source an on-sky calibrations to ensures robust NCPA correction, with future work focusing on optimizing DM2 stroke requirements.

\acknowledgments 
The SAXO\texttt{+} instrument is designed and being built by a consortium consisting of LIRA, CRAL, INAF, IPAG, Lagrange, Institut d'Optique, University of Geneva, MPIA, and LAM, in collaboration with ESO. The SAXO\texttt{+} project is funded by the Fondation Charles Defforey-Institut de France, CNRS/INSU, the Programme et Equipements Prioritaires de Recherche PEPR Origins (ANR-22-EXOR-0002, ANR-22-EXOR-0003, ANR-22-EXOR-0017)  within the Plan France 2030 of the French government operated by the National Research Agency (ANR), the NextGenerationEU funds within the National Recovery and Resilience Plan (PNRR) - M4C2 Inv. 3.1 - Project IR0000034 "STILES" - CUP C33C22000640006, PlanetS, and MPIA. 

\noindent \textit{Software} - This study used the following python packages:
\texttt{Asterix}\cite{mazoyer_AsterixSimulator}, 
\texttt{Astropy}\cite{astropy_collaboration_2022},
\texttt{COMPASS}\cite{gratadour2016_COMPASSStatusUpdate},
\texttt{Matplotlib} \cite{hunter2007matplotlib},
\texttt{NumPy}\cite{harris2020NumPy}, \texttt{pyZELDA}\cite{Vigan18_pyZELDA} and
\texttt{SciPy}\cite{virtanen2020scipy}.

\bibliography{report} 

@article{vigan2019_CalibrationQuasistaticAberrations,
  title = {Calibration of Quasi-Static Aberrations in Exoplanet Direct-Imaging Instruments with a {{Zernike}} Phase-Mask Sensor. {{III}}. {{On-sky}} Validation in {{VLT}}/{{SPHERE}}},
  author = {Vigan, A. and N'Diaye, M. and Dohlen, K. and Sauvage, J.-F. and Milli, J. and Zins, G. and Petit, C. and Wahhaj, Z. and Cantalloube, F. and Caillat, A. and Costille, A. and Le Merrer, J. and Carlotti, A. and Beuzit, J.-L. and Mouillet, D.},
  year = {2019},
  month = sep,
  journal = {Astronomy and Astrophysics},
  volume = {629},
  pages = {A11},
  issn = {0004-6361},
  doi = {10.1051/0004-6361/201935889},
  urldate = {2022-03-11},
  langid = {english}
}

@inproceedings{mazoyer_upgrading_2024,
	title = {Upgrading {SPHERE} with the second stage {AO} system {SAXO}+: non-common path aberrations estimation and correction},
	volume = {13096},
	url = {https://www.spiedigitallibrary.org/conference-proceedings-of-spie/13096/130969D/Upgrading-SPHERE-with-the-second-stage-AO-system-SAXO/10.1117/12.3019921.full},
	doi = {10.1117/12.3019921},
	shorttitle = {Upgrading {SPHERE} with the second stage {AO} system {SAXO}+},
	eventtitle = {Ground-based and Airborne Instrumentation for Astronomy X},
	pages = {2747--2762},
	booktitle = {Proceedings of the {SPIE}},
	publisher = {{SPIE}},
	author = {Mazoyer, Johan and Goulas, Charles and Vidal, Fabrice and Dinis, Isaac Bernardino and Milli, Julien and Tallon, Michel and Galicher, Raphaël and Absil, Olivier and Bechet, Clémentine and Boccaletti, Anthony and Ferreira, Florian and Langlois, Maud and Martinez, Patrice and Mugnier, Laurent and N'diaye, Mamadou and Xivry, Gilles Orban de and Potier, Axel and Tallon-Bosc, Isabelle and Vigan, Arthur},
	urldate = {2024-09-02},
	date = {2024-07-18},
	year = {2024},
	month = {07},
}

@article{goulas_non-common_2026,
	title = {Non-common path aberration compensation and a dark hole loop with a pyramid adaptive optics system: Application to {SAXO}+},
	volume = {708},
	rights = {© The Authors 2026},
	issn = {0004-6361, 1432-0746},
	url = {https://www.aanda.org/articles/aa/abs/2026/04/aa56168-25/aa56168-25.html},
	doi = {10.1051/0004-6361/202556168},
	shorttitle = {Non-common path aberration compensation and a dark hole loop with a pyramid adaptive optics system},
	pages = {A50},
	journal = {Astronomy \& Astrophysics},
	shortjournal = {A\&A},
	publisher = {{EDP} Sciences},
	author = {Goulas, C. and Galicher, R. and Vidal, F. and Mazoyer, J. and Ferreira, F. and Sevin, A. and Potier, A. and Boccaletti, A. and Gendron, E. and Béchet, C. and Tallon, M. and Langlois, M. and Kulcsár, C. and Raynaud, H.-F. and Galland, N. and Schreiber, L. and Dinis, I. Bernardino and Wildi, F. and Chauvin, G.},
	urldate = {2026-06-09},
	date = {2026-04-01},
	year = {2026},
	month = {04},
}

@INPROCEEDINGS{Males2018SPIE_MagAO-X,
       author = {{Males}, Jared R. and {Close}, Laird M. and others},
        title = "{MagAO-X: project status and first laboratory results}",
    booktitle = {Adaptive Optics Systems VI},
         year = 2018,
       editor = {{Close}, Laird M. and {Schreiber}, Laura and {Schmidt}, Dirk},
       series = {Society of Photo-Optical Instrumentation Engineers (SPIE) Conference Series},
       volume = {10703},
        month = jul,
          eid = {1070309},
        pages = {1070309},
          doi = {10.1117/12.2312992},
archivePrefix = {arXiv},
       eprint = {1807.04315},
 primaryClass = {astro-ph.IM},
       adsurl = {https://ui.adsabs.harvard.edu/abs/2018SPIE10703E..09M}
}

@article{vigan2022_CalibrationQuasistaticAberrations,
  title = {Calibration of Quasi-Static Aberrations in Exoplanet Direct-Imaging Instruments with a {{Zernike}} Phase-Mask Sensor. {{IV}}. {{Temporal}} Stability of Non-Common Path Aberrations in {{VLT}}/{{SPHERE}}},
  author = {Vigan, A. and Dohlen, K. and N'Diaye, M. and Cantalloube, F. and Girard, J. H. and Milli, J. and Sauvage, J.-F. and Wahhaj, Z. and Zins, G. and Beuzit, J.-L. and Caillat, A. and Costille, A. and Le Merrer, J. and Mouillet, D. and Tourenq, S.},
  year = {2022},
  month = apr,
  journal = {Astronomy and Astrophysics},
  volume = {660},
  pages = {A140},
  issn = {0004-6361},
  doi = {10.1051/0004-6361/202142635},
  urldate = {2022-06-14},
  langid = {english}
}

@INPROCEEDINGS{Potier2022spie,
       author = {{Potier}, Axel and {Wahhaj}, Zahed and {Galicher}, Raphael and {Mazoyer}, Johan and {Baudoz}, Pierre and {Chauvin}, Gael and {Ruane}, Garreth},
        title = "{Improving VLT/SPHERE without additional hardware: comparing quasi-static correction strategies}",
    booktitle = {Adaptive Optics Systems VIII},
         year = 2022,
       editor = {{Schreiber}, Laura and {Schmidt}, Dirk and {Vernet}, Elise},
       series = {Society of Photo-Optical Instrumentation Engineers (SPIE) Conference Series},
       volume = {12185},
        month = aug,
          eid = {1218568},
        pages = {1218568},
          doi = {10.1117/12.2629926},
archivePrefix = {arXiv},
       eprint = {2305.19501},
 primaryClass = {astro-ph.IM},
       adsurl = {https://ui.adsabs.harvard.edu/abs/2022SPIE12185E..68P}
}

@software{Vigan18_pyZELDA,
       author = {{Vigan}, A. and {N'Diaye}, M.},
        title = "{pyZELDA: Python code for Zernike wavefront sensors}",
 howpublished = {Astrophysics Source Code Library, record ascl:1806.003},
         year = 2018,
        month = jun,
          eid = {ascl:1806.003},
       adsurl = {https://ui.adsabs.harvard.edu/abs/2018ascl.soft06003V}
}

@article{esposito2020_OnskyCorrectionNoncommon,
  title = {On-Sky Correction of Non-Common Path Aberration with the Pyramid Wavefront Sensor},
  author = {Esposito, S. and Puglisi, A. and Pinna, E. and Agapito, G. and {Quir{\'o}s-Pacheco}, F. and V{\'e}ran, J. P. and Herriot, G.},
  year = {2020},
  month = apr,
  journal = {Astronomy \& Astrophysics},
  volume = {636},
  pages = {A88},
  publisher = {EDP Sciences},
  issn = {0004-6361, 1432-0746},
  doi = {10.1051/0004-6361/201937033},
  urldate = {2023-10-02},
  copyright = {{\copyright} ESO 2020},
  langid = {english}
}

@article{beuzit2019_SPHEREExoplanetImager,
  title = {{{SPHERE}}: The Exoplanet Imager for the {{Very Large Telescope}}},
  shorttitle = {{{SPHERE}}},
  author = {Beuzit, J.-L. and Vigan, A. and Mouillet, D. and others},
  year = {2019},
  month = nov,
  journal = {Astronomy and Astrophysics},
  volume = {631},
  pages = {A155},
  issn = {0004-6361},
  doi = {10.1051/0004-6361/201935251},
  urldate = {2020-11-24}
}

@inproceedings{lozi2020_StatusSCExAOInstrument,
  title = {Status of the {{SCExAO}} Instrument: Recent Technology Upgrades and Path to a System-Level Demonstrator for {{PSI}}},
  shorttitle = {Status of the {{SCExAO}} Instrument},
  booktitle = {Proceedings of the {{SPIE}}},
  author = {Lozi, Julien and Guyon, Olivier and Vievard, S{\'e}bastien and others},
  year = {2020},
  month = dec,
  volume = {11448},
  pages = {110--121},
  publisher = {SPIE},
  doi = {10.1117/12.2562832},
  urldate = {2023-09-26}
}

@article{macintosh2014_FirstLightGemini,
  title = {First Light of the {{Gemini Planet Imager}}},
  author = {Macintosh, Bruce and Graham, James R. and Ingraham, Patrick and others},
  year = {2014},
  month = sep,
  journal = {Proceedings of the National Academy of Sciences},
  volume = {111},
  number = {35},
  pages = {12661--12666},
  publisher = {National Academy of Sciences},
  issn = {0027-8424, 1091-6490},
  doi = {10.1073/pnas.1304215111},
  urldate = {2020-11-24},
  chapter = {Physical Sciences},
  langid = {english},
  pmid = {24821792}
}

@article{korkiakoski2008_ImprovingPerformancePyramid,
  title = {Improving the Performance of a Pyramid Wavefront Sensor with Modal Sensitivity Compensation},
  author = {Korkiakoski, Visa and V{\'e}rinaud, Christophe and Louarn, Miska Le},
  year = {2008},
  month = jan,
  journal = {Applied Optics},
  volume = {47},
  pages = {79},
  issn = {0003-6935},
  doi = {10.1364/AO.47.000079},
  urldate = {2024-03-17}
}

@inproceedings{gratadour2016_COMPASSStatusUpdate,
  title = {{{COMPASS}}: Status Update and Long Term Development Plan},
  shorttitle = {{{COMPASS}}},
  booktitle = {Adaptive {{Optics Systems V}}},
  author = {Gratadour, D. and Ferreira, F. and Sevin, A. and Doucet, N. and Cl{\'e}net, Y. and Gendron, E. and Lain{\'e}, M. and Vidal, F. and Brul{\'e}, J. and Puech, M. and V{\'e}rinaud, C. and Carlotti, A.},
  year = {2016},
  month = jul,
  volume = {9909},
  pages = {2057--2063},
  publisher = {SPIE},
  doi = {10.1117/12.2232681},
  urldate = {2024-06-19}
}

@article{ndiaye2016_ApodizedPupilLyot,
  title = {Apodized {{Pupil Lyot Coronagraphs}} for {{Arbitrary Apertures}}. {{V}}. {{Hybrid Shaped Pupil Designs}} for {{Imaging Earth-like}} Planets with {{Future Space Observatories}}},
  author = {N'Diaye, Mamadou and Soummer, R{\'e}mi and Pueyo, Laurent and Carlotti, Alexis and Stark, Christopher C. and Perrin, Marshall D.},
  year = {2016},
  month = feb,
  journal = {The Astrophysical Journal},
  volume = {818},
  pages = {163},
  issn = {0004-637X},
  doi = {10.3847/0004-637X/818/2/163},
  urldate = {2016-06-17}
}

@online{mazoyer_AsterixSimulator,
	title = {Asterix: Simulate your High-Contrast Instruments},
	url = {https://asterix-hci.readthedocs.io/},
	author = {Mazoyer, J. and Potier, A. and Laginja, I. and Galicher, R.},
	urldate = {2024-01-05},
	year = {2019},
    note = {\url{https://asterix-hci.readthedocs.io/}},
}

@article{astropy_collaboration_2022,
    title = {The Astropy Project: Sustaining and Growing a Community-oriented Open-source Project and the Latest Major Release (v5.0) of the Core Package},
	volume = {935},
	issn = {0004-637X},
	url = {https://ui.adsabs.harvard.edu/abs/2022ApJ...935..167A},
	doi = {10.3847/1538-4357/ac7c74},
	shorttitle = {The Astropy Project},
	pages = {167},
	journal = {The Astrophysical Journal},
	publisher = {{IOP}},
	author = {{Astropy Collaboration et al.}},	
	urldate = {2026-06-22},
	date = {2022-08-01},
	year = {2022},
	month = {08},
}

@inproceedings{tallon2026_SAXOpluscontrol,
author = {Tallon, Michel and Béchet, Clémentine and Dinis, Isaac and
Ferreira, Florian and Fontanillas, Paul and Galland, Nicolas and Goulas, Charles and  Kasper, Markus and Kulcs{'a}r, Caroline  and Nousiainen, Jalo and Raynaud, Henri-Fran{\c c}ois and Tallon-Bosc, Isabelle and Thiébaut, {'E}ric and Langlois, Maud and  Boccaletti, Anthony and Diolaiti, Emiliano and Galicher, Raphaël and Lelouarn, Miska and Loupias, Magali and Mazoyer, Johan  and  Milli, Julien and  Schreiber, Laura  and  Vidal, Fabrice and BArbary, Gaël and Wildi, Fran{\c c}ois},
eventtitle = {Adaptive Optics Systems X},
title = {SAXO+, the second-stage adaptive optics for SPHERE: a review of six selected control algorithms and the results achieved to date},
booktitle = {Proceedings of SPIE Astronomical Telescopes + Instrumentation},
year = {2026},
publisher = {{SPIE}},
pages = {14150-38},
note = {in these proceedings}}

@inproceedings{Langlois2026_SAXOplussimu,
author = {Langlois, Maud and Béchet, Clémentine and Schreiber, Laura and Tallon, Michel and Ferreira, Florian and Milli, Julien and Loupias, Magali and Dinis, Isaac and Diolaiti, Emiliano and Feldt, Markus and Galicher, Raphaël and Kulcsár, Caroline and Mazoyer, Johan and N'Diaye, Mamadou and Raynaud, Henri-François and Goulas, Charles and Stadler, Eric and Tallon-Bosc, Isabelle and Boccaletti, Anthony and Chauvin, Gaël and Galland, Nicolas and Lelouarn, Miska and Kasper, Markus and Nousiainen, Jalo and Vigan, Arthur and Thiébaut, {'E}ric and Vidal, Fabrice and Potier, Axel and Wildi, Fran{\c c}ois},
eventtitle = {Adaptive Optics Systems X},
title = {SAXO+, the second-stage adaptive optics for SPHERE: high contrast performance analysis},
booktitle = {Proceedings of SPIE Astronomical Telescopes + Instrumentation},
year = {2026},
publisher = {SPIE},
pages = {14150-122},
note = {in these proceedings}
}

@article{giveon_closed_2007,
	title = {Closed loop, {DM} diversity-based, wavefront correction algorithm for high contrast imaging systems},
	volume = {15},
	issn = {1094-4087},
	url = {https://ui.adsabs.harvard.edu/abs/2007OExpr..1512338G},
	doi = {10.1364/OE.15.012338},
	pages = {12338},
	journal = {Optics Express},
	author = {Give'on, Amir and Belikov, Ruslan and Shaklan, Stuart and Kasdin, Jeremy},
	urldate = {2025-01-09},
	date = {2007-09-01},
	year = {2007},
	month = {09},
	note = {{ADS} Bibcode: 2007OExpr..1512338G},
}

@article{potier_increasing_2022,
	title = {Increasing the raw contrast of {VLT}/{SPHERE} with the dark hole technique. {II}. On-sky wavefront correction and coherent differential imaging},
	volume = {665},
	issn = {0004-6361},
	url = {https://ui.adsabs.harvard.edu/abs/2022A&A...665A.136P/abstract},
	doi = {10.1051/0004-6361/202244185},
	pages = {A136},
	journal = {Astronomy and Astrophysics},
	shortjournal = {A\&A},
	author = {Potier, A. and Mazoyer, J. and Wahhaj, Z. and Baudoz, P. and Chauvin, G. and Galicher, R. and Ruane, G.},
	urldate = {2022-10-27},
	date = {2022-09},
	langid = {english},
  year = {2022},
  month = {sep},
}

@article{galicher_imaging_2024,
	title = {Imaging exoplanets with coronagraphic instruments},
	volume = {24},
	issn = {1631-0705},
	url = {https://ui.adsabs.harvard.edu/abs/2024CRPhy..24S.133G},
	doi = {10.5802/crphys.133},
	pages = {133},
    journal = {Comptes Rendus Physique},
    shortjournal = {C. R. Phys.},
	author = {Galicher, Raphaël and Mazoyer, Johan},
	urldate = {2024-06-19},
	date = {2024-01-01},
	note = {{ADS} Bibcode: 2024CRPhy..24S.133G},
  year = {2024},
  month = {jan},
}

@article{boccaletti_sphereplus_2020,
	title = {{SPHERE}+: Imaging young Jupiters down to the snowline},
	volume = {2003},
	url = {http://adsabs.harvard.edu/abs/2020arXiv200305714B},
	doi = {10.48550/arXiv.2003.05714},
	pages = {arXiv:2003.05714},
	journal = {{arXiv} e-prints},
	shortjournal = {{arXiv} e-prints},
	author = {Boccaletti, A. and Chauvin, G. and Mouillet, D. and others},
	urldate = {2020-05-05},
	date = {2020-03-01},
	year = {2020},
}

@article{cantalloube_sphere_2019,
	title = {Peering through {SPHERE} Images: A Glance at Contrast Limitations},
	volume = {176},
	issn = {0722-6691},
	url = {https://ui.adsabs.harvard.edu/abs/2019Msngr.176...25C},
	doi = {10.18727/0722-6691/5138},
	shorttitle = {Peering through {SPHERE} Images},
	pages = {25--31},
	journal = {The Messenger},
	shortjournal = {The Messenger},
	author = {Cantalloube, F. and Dohlen, K. and Milli, J. and Brandner, W. and Vigan, A.},
	urldate = {2021-06-17},
	date = {2019-06-01},
	year = {2019},
}

@article{deo2019_TelescopereadyApproachModala,
  title = {A Telescope-Ready Approach for Modal Compensation of Pyramid Wavefront Sensor Optical Gain},
  author = {Deo, V. and Gendron, {\'E} and Rousset, G. and Vidal, F. and Sevin, A. and Ferreira, F. and Gratadour, D. and Buey, T.},
  year = {2019},
  month = sep,
  journal = {Astronomy \& Astrophysics},
  volume = {629},
  pages = {A107},
  publisher = {EDP Sciences},
  issn = {0004-6361, 1432-0746},
  doi = {10.1051/0004-6361/201935847},
  urldate = {2024-06-24},
  langid = {english}
}

@article{chambouleyron2020_PyramidWavefrontSensor,
  title = {Pyramid Wavefront Sensor Optical Gains Compensation Using a Convolutional Model},
  author = {Chambouleyron, V. and Fauvarque, O. and {Janin-Potiron}, P. and Correia, C. and Sauvage, J.-F. and Schwartz, N. and Neichel, B. and Fusco, T.},
  year = {2020},
  month = dec,
  journal = {Astronomy \& Astrophysics},
  volume = {644},
  pages = {A6},
  publisher = {EDP Sciences},
  issn = {0004-6361, 1432-0746},
  doi = {10.1051/0004-6361/202037836},
  urldate = {2024-04-11},
  langid = {english}
}

@ARTICLE{Vigan_zelda_2019,
       author = {{Vigan}, A. and {N'Diaye}, M. and {Dohlen}, K. and {Sauvage}, J. -F. and {Milli}, J. and {Zins}, G. and {Petit}, C. and {Wahhaj}, Z. and {Cantalloube}, F. and {Caillat}, A. and {Costille}, A. and {Le Merrer}, J. and {Carlotti}, A. and {Beuzit}, J. -L. and {Mouillet}, D.},
        title = "{Calibration of quasi-static aberrations in exoplanet direct-imaging instruments with a Zernike phase-mask sensor. III. On-sky validation in VLT/SPHERE}",
      journal = {Astronomy and Astrophysics},
         year = 2019,
        month = sep,
       volume = {629},
          eid = {A11},
        pages = {A11},
          doi = {10.1051/0004-6361/201935889},
archivePrefix = {arXiv},
       eprint = {1907.11241},
 primaryClass = {astro-ph.IM},
       adsurl = {https://ui.adsabs.harvard.edu/abs/2019A&A...629A..11V}
}

@ARTICLE{Ragazzoni_1996,
       author = {{Ragazzoni}, Roberto},
        title = "{Pupil plane wavefront sensing with an oscillating prism}",
      journal = {Journal of Modern Optics},
         year = 1996,
        month = feb,
       volume = {43},
       number = {2},
        pages = {289-293},
          doi = {10.1080/09500349608232742},
       adsurl = {https://ui.adsabs.harvard.edu/abs/1996JMOp...43..289R}
}

@article{harris2020NumPy,
  title={Array programming with {NumPy}},
  author={Harris, Charles R. and Millman, K. Jarrod and van der Walt, Stéfan J. and Gommers, Ralf and Virtanen, Pauli and Cournapeau, David and Wieser, Eric and Taylor, Julian and Berg, Sebastian and Smith, Nathaniel J. and Kern, Robert and Picus, Matt and Hoyer, Stephan and van Kerkwijk, Marten H. and others},
  journal={Nature},
  volume={585},
  number={7825},
  pages={357--362},
  year={2020},
  publisher={Nature Publishing Group},
  doi={10.1038/s41586-020-2649-2}
}

@article{hunter2007matplotlib,
  title={Matplotlib: A 2D graphics environment},
  author={Hunter, John D.},
  journal={Computing in Science \& Engineering},
  volume={9},
  number={3},
  pages={90--95},
  year={2007},
  publisher={IEEE},
  doi={10.1109/MCSE.2007.55}
}

@article{virtanen2020scipy,
  title={SciPy 1.0: Fundamental Algorithms for Scientific Computing in Python},
  author={Virtanen, Pauli and Gommers, Ralf and Oliphant, Travis E. and Haberland, Matt and Reddy, Tyler and Cournapeau, David and Burovski, Evgeni and Peterson, Pearu and Weckesser, Warren and Bright, Jonathan and van der Walt, Stéfan J. and Brett, Matthew and Wilson, Joshua and Millman, K. Jarrod and others},
  journal={Nature Methods},
  volume={17},
  number={3},
  pages={261--272},
  year={2020},
  publisher={Nature Publishing Group},
  doi={10.1038/s41592-019-0686-2}
}

@ARTICLE{Squicciarini_SHINE_2026,
       author = {{Squicciarini}, V. and {Desidera}, S. and {Chauvin}, G. and others},
        title = "{The SPHERE infrared survey for exoplanets (SHINE) V. Full sample characterization}",
      journal = {arXiv e-prints},
         year = 2026,
        month = may,
          eid = {arXiv:2605.27247},
        pages = {arXiv:2605.27247},
          doi = {10.48550/arXiv.2605.27247},
archivePrefix = {arXiv},
        volume = { a},
       eprint = {2605.27247},
 Notes = {Accepted in A\&A}
}
\bibliographystyle{spiebib} 

\end{document}